\documentclass{PoS}
\usepackage{graphicx}
\usepackage{wrapfig}
\usepackage{amsmath}   
\usepackage{times}
\newcommand{\bea}{\begin{eqnarray}}
\newcommand{\eea}{\end{eqnarray}}
\newcommand{\bqa}{\begin{eqnarray}}
\newcommand{\eqa}{\end{eqnarray}}

\newcommand{\eps}{\varepsilon}
\newcommand{\bq}{ \begin {equation} }
\newcommand{\eq}{\end{equation}}
\newcommand{\be}{\begin{eqnarray}}
\newcommand{\ea}{\end{eqnarray}}

\title{%
{\small DESY 10--012 \newline SFB/CPP--10--16 \newline HEPTOOLS 10--073} \\[1.5cm]
A recursive approach to the reduction of tensor Feynman integrals}

\ShortTitle{A recursive approach to the reduction of tensor Feynman integrals}

{\author{Theodoros Diakonidis\\
Deutsches Elektronen-Synchrotron, DESY, Platanenallee 6, 15738 Zeuthen, Germany
\\
        E-mail: \email{Theodoros.Diakonidis@desy.de}}}

\author{Jochem Fleischer
\\
Fakult\"at f\"ur Physik, Universit\"at Bielefeld, Universit\"atsstr. 25,  33615
Bielefeld, Germany
\\
        E-mail: \email{Fleischer@physik.uni-bielefeld.de}}

\author{\speaker{Tord Riemann}
        \\
        Deutsches Elektronen-Synchrotron, DESY, Platanenallee 6, 15738 Zeuthen, Germany
        \\
        E-mail: \email{Tord.Riemann@desy.de}}

\author{Bas Tausk
\\
Deutsches Elektronen-Synchrotron, DESY, Platanenallee 6, 15738 Zeuthen, Germany
\\
        E-mail: \email{Bas.Tausk@desy.de}}

\abstract{%
We describe a new, convenient, recursive tensor integral reduction scheme for one-loop $n$-point Feynman integrals.
The reduction is based on the algebraic Davydychev-Tarasov formalism where the tensors are represented by scalars with shifted dimensions and indices, and then expressed by conventional scalars with generalized recurrence relations.
The scheme is worked out explicitly for up to $n=6$ external legs and for tensor ranks $R\leq n$.
The tensors are represented by scalar one- to four-point functions in $d$ dimensions.
For the evaluation of them, the Fortran code  for the tensor reductions has to be linked with a package like QCDloop or LoopTools/FF.
Typical numerical results are presented.
}

\FullConference{RADCOR 2009 - 9th International Symposium on Radiative Corrections (Applications of Quantum Field Theory to Phenomenology) ,\\
		 October 25 - 30 2009\\
		 Ascona, Switzerland}

\begin{document}
\section{Introduction}
$n$-point integrals $I_n^R$ in loop momentum space of tensor rank $R$ appear in any realistic evaluation of Feynman diagrams.
There are several ways to calculate them, and one approach expresses them by a small set of scalar integrals.
The first systematic treatment, in a Standard Model calculation, is known as the Passarino-Veltman reduction \cite{Passarino:1978jh} and expresses four-point tensor integrals (and simpler ones) algebraically by scalar one- to four-point functions.
Two of us made use of this scheme in the early days \cite{Fleischer:1980ub,Riemann:1981ga,Mann:1983dv}.
Nowadays tensor reductions became again a topic of research because at LEP2, LHC and ILC the interesting final states typically consist of more than two particles, some of them being massive.
Several dedicated tensor reduction packages  have been developed for the calculation of  five- and six-point functions.
As open-source packages we like to mention the Fortran packages LoopTools/FF  \cite{Hahn:1998yk2,vanOldenborgh:1990yc} (covering  $I_n^R$ with $n \leq 5,R \leq 4$) and Golem95 \cite{Binoth:2008uq}  (covering  $I_n^R$ with $n\leq 6$ and  massless propagators) and the Mathematica package hexagon.m \cite{Diakonidis:2008dt,Diakonidis:2008ij} (covering  $I_n^R$ with $(n,R) \leq (6,4), (5,3)$).

Here, we describe a recursive implementation for tensor functions $I_n^R$ with $n\leq 6, R\leq n$ for arbitrary internal masses.
We use the Davydychev-Tarasov approach where the tensor integrals are first expressed by scalar integrals with higher dimensions and indices \cite{Davydychev:1991va}.
In a second step, the scalar integrals may be expressed algebraically by   scalar one- to four-point functions \cite{Tarasov:1996br} quite similar to the Passarino-Veltman reduction.
In fact, when using the same basis, the approaches are equivalent.
A difference, though, may arise in the algorithmic realization, and as a consequence in the numerical stability and speed of an implementation.
For more comments on the differences of tensor reduction schemes we refer to the literature quoted.
In a recent paper \cite{Diakonidis:2009fx}, we introduced a convenient and easy-to-program version of the reduction a la Davydychev-Tarasov, which allows a recursive determination of a chain of tensors.
Here we will describe that scheme and present some numerical results.
The integrals to be evaluated are:
\begin{eqnarray}\label{definition}
 I_n^{\mu_1\cdots\mu_R} &=&  ~ C(\varepsilon) ~\int \frac{d^d k}{i\pi^{d/2}}~~\frac{\prod_{r=1}^{R} k^{\mu_r}}{\prod_{j=1}^{n}c_j^{\nu_j}},
\end{eqnarray}
where the denominators $c_j$ have \emph{indices} $\nu_j$ and \emph{chords}
$q_j$:
\begin{eqnarray}\label{propagators}
c_j &=& (k-q_j)^2-m_j^2 +i \epsilon .
\end{eqnarray}
The normalization $C(\varepsilon)$ plays a role for divergent integrals only and is conventional:
\begin{eqnarray}
C(\varepsilon) &=&  (\mu)^{2\eps}~\frac{\Gamma(1 - 2\eps)}{\Gamma(1 + \eps) \Gamma^2(1 - \eps)}.
\end{eqnarray}
Here, we use $d=4-2\epsilon$ and $\mu=1$.
For the evaluation of the scalar functions we will rely on either LoopTools or QCDloops/FF \cite{Ellis:2007qk,vanOldenborgh:1990yc}, and the latter one uses also $C(\varepsilon)$ as defined here.

\section{Recursions}
Our recursions begin with six-point functions where the well known formula \cite{Fleischer:1999hq,Binoth:2005ff,Denner:2005nn,Diakonidis:2008ij} may be used:
\bea\label{tensor6general}
I_6^{\mu_1 \dots \mu_{R-1} \rho}  =
-  \sum_{s=1}^{6}
I_5^{\mu_1 \dots \mu_{R-1} ,s } \bar{Q}_s^{\rho}.
\eea
The auxiliary vectors $\bar{Q}_s^{\rho}$ read:
\bea
 \bar{Q}_s^{\rho}&=&\sum_{i=1}^{6}  q_i^{\rho} \frac{{0s\choose 0i}_6}{{0\choose 0}_6}~~~,~~~ s=1 \dots 6.
\label{Q6}
\eea
The $I_{n-1}^{\{\mu_1,\cdots\},s}$ is obtained from $I_{n}^{\{\mu_1,\cdots\}}$ by shrinking line $s$, and the ${i,j,\cdots \choose k,l,\cdots}_n$ are signed minors of the modified Cayley determinant ${\choose}_n$ \cite{Melrose:1965kb}.
For further details of notations we refer to \cite{Diakonidis:2009fx}.

\begin{figure}[tb]
\begin{center}
\rotatebox{-90}{\includegraphics*[angle=0,height=15.0cm]{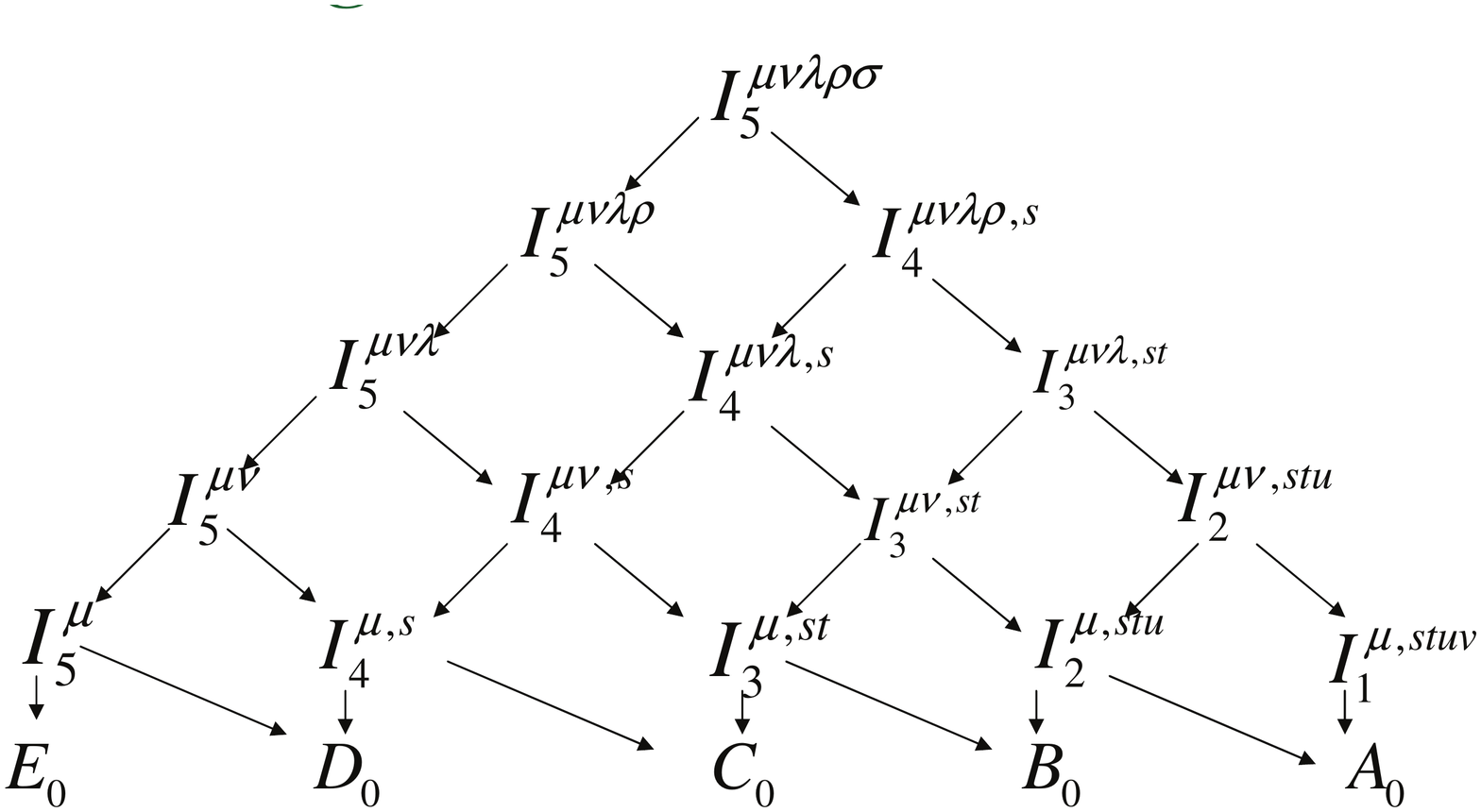}}
\caption{\label{fl-triangle}The recursion triangle.}
\end{center}\end{figure}

The further calculational chain may be read off from figure \ref{fl-triangle}.
Its basic idea is to represent an $n$-point tensor $I_n^R$ by an $n$-point tensor of lower rank $I_n^{R-1}$ and by all the $(n-1)$-point tensors of lower rank  $I_{n-1}^{R-1}$.
\footnote{Similar recursive realizations of the Passarino-Veltman reduction may be found in \cite{Denner:2005nn}, see there figures~ 2 and~ 3.}
For 5-point functions, we derived in \cite{Diakonidis:2009fx}:
\bea\label{tensor5general}
I_5^{\mu_1  \dots \mu_{R-1} \mu}  &=&I_5^{\mu_1  \dots \mu_{R-1}} Q_0^{\mu} -  \sum_{s=1}^{5}
I_4^{\mu_1  \dots \mu_{R-1},s } Q_s^{\mu},
\\
\label{Qs}
 Q_s^{\mu}&=&\sum_{i=1}^{n}  q_i^{\mu} \frac{{s\choose i}_n}{\left(  \right)_n},~~~ s=0, \dots, n.
\eea
The formula is the analogue to (\ref{tensor6general}).
For $n$-point functions with $n<5$, the corresponding representations contain additional terms because the number of independent chords is then less than four so that the chords don't form a complete basis for $d=4$.
There are several modifications to be applied, and we like to reproduce only one example with auxiliary terms:
\bea
\label{tensor44}
I_4^{\mu \nu \lambda \rho}  &=&I_4^{\mu \nu \lambda} Q_0^{\rho} -  \sum_{t=1}^{4}
I_3^{\mu \nu \lambda,t } Q_t^{\rho}
-
G^{\mu \rho } T^{\nu \lambda }- G^{\nu \rho } T^{\mu \lambda }-G^{\lambda \rho } T^{\mu \nu },
\eea
with the additional tensor and vector  components:
\bea \label{eq03}
T^{\mu \nu }&=&I_4^{\mu,[d+]} Q_0^{\nu} - \sum_{t=1}^{4} I_3^{\mu,[d+],t} ~ Q_t^{\nu}~-G^{\mu \nu }I_{4}^{[d+]^2} ,
\\
\label{Gml}
G^{\mu \lambda}&=&\frac{1}{2} g^{\mu \lambda}-
\sum_{i,j=1}^{4} q_i^{\mu}  q_j^{\lambda} \frac{{i\choose j}_4}{\left(  \right)_4},
\\
I_4^{\mu,[d+]}
&=& I_4^{[d+]}
Q_0^{\mu} - \sum_{t=1}^{4} I_3^{[d+],t} Q_t^{\mu},
\label{GV}
\\
I_3^{\mu,[d+],t}
&=&
I_3^{[d+],t} Q_0^{t,\mu} - \sum_{u=1}^{4} I_2^{[d+],tu} Q_u^{t,\mu},
\label{Wt}
\\\label{Qst}
 Q_u^{t,\mu}&=&\sum_{i=1}^{4}  q_i^{\mu} \frac{{ut\choose it}_4}{ {t \choose t}_4},~~~ u=0, \dots, 4.
\eea
They may be, finally, represented by the scalar integrals in $d$ dimensions $I_{4},I_{3  }^{t},I_{2  }^{tu}, I_{1}^{tuw}$, where the indices $t,u,w$ indicate truncations of corresponding lines:
\bea
\label{eq04}
I_{4}^{[d+]^2}&=&\left[\frac{{0\choose 0}_4}{\left(  \right)_4} I_{4}^{[d+]} -
 \sum_{t=1}^{4} \frac{{t\choose 0}_4}{\left(  \right)_4} I_{3}^{[d+],t} \right]{\frac{1}{d-1}},
\\
I_{4}^{[d+]}&=& \frac{{0\choose 0}_4}{{\choose }_4} I_{4}
 -
  \sum_{t=1}^{4} \frac{{t\choose 0}_4}{{\choose }_4} I_{3}^{t},
\\
 I_{3  }^{[d+],t}&=& \left[
\frac{{0t\choose 0t}_4}{{t\choose t}_4}I_{3  }^{t}-
\sum_{u=1}^4\frac{ {ut\choose 0t}_4}{{t\choose t}_4} I_{2  }^{tu} \right]{\frac{1}{d-2}} ,
\label{A301}
\\
\label{eq05}
I_{2}^{[d+],tu}&=&\left[\frac{{0tu\choose 0tu}_4}{{tu\choose tu}_4} I_{2}^{tu}-
 \sum_{w=1}^{4} \frac{{0tu\choose wtu}_4}{{tu\choose tu}_4} I_{1}^{tuw} \right]{\frac{1}{d-1}}
.
\eea
The representations  for the simpler tensors have been given in \cite{Diakonidis:2009fx}.

\section{Numerical results}

\begin{table}[tb]
\centering
\begin{tabular}{|l|r@{.}l|r@{.}l|r@{.}l|r@{.}l|}
\hline
$p_1$ &  0&5         &  0&0         &  0&0         &  0&5        \\
$p_2$ &  0&5         &  0&0         &  0&0         &-- 0&5        \\
$p_3$ & -- 0&19178191&-- 0&12741180  &-- 0&08262477  &-- 0&11713105  \\
$p_4$ & -- 0&33662712 &  0&06648281  &  0&31893785  &  0&08471424  \\
$p_5$ & -- 0&21604814 &  0&20363139  &-- 0&04415762  &-- 0&05710657
\\ \hline
\multicolumn{9}{|c|}{$p_6 =-(p_1+p_2+p_3+p_4+p_5)$}
\\
\hline
\end{tabular}
\caption{\label{kinem-hexagon-massless}Phase space point of  massless six-point functions taken from \cite{Binoth:2008uq}. }
\end{table}

An example of a completely massless tensor reduction uses the momenta given in table \ref{kinem-hexagon-massless}.
We combined our tensor reduction with the scalar master integrals from QCDloop \cite{Ellis:2007qk}.
Table  \ref{sixpoint-golem}  contains sample tensor components of a six-point function with rank $R=5$.
It  shows an agreement of eight digits between the results of our package Hexagon.F \cite{diakonidis-hexagonV0.9:2009} and those of Golem95 \cite{Binoth:2008uq} for the constant terms of the tensor components.

\begin{table}[htb]
\begin{center}
\begin{tabular}{|l|r@{.}l r@{.}l  |  r@{.}l r@{.}l     |}
\hline
          & \multicolumn{4}{|c|}{Hexagon.F} & \multicolumn{4}{|c|}{Golem95}
\\
\hline
${F^{03121}}$ & 0&158428987E+0,& 0&41670698E--1   &  0&158428981E+0 ,& 0&41670700E--1
\\
\hline
${F^{11020}}$
& -- 0 & 143913860E+1,& -- 0&16464705E+0 & -- 0&143913853E+1 ,& -- 0&16464708E+0\\
\hline
${F^{20200}}$
& 0&242928780E+2,& 0&55504184E+2   & 0&242928776E+2 ,&  0&55504182E+2
\\
\hline
${F^{22130}}$
& 0&225563941E+0,& 0&23192857E+0  & 0&225563949E+0 ,&  0&23192851E+0
\\
\hline
${F^{33333}}$
& 0&244568135E+0,& 0&74014604E+0   & 0&244568138E+0 ,& 0&74014610E+0\\
 \hline
\end{tabular}
\end{center}
\caption[]{\label{sixpoint-golem}Real and imaginary parts of selected tensor components of rank $R=5$ massless hexagon integrals;
comparison of the packages Hexagon.F and Golem95.}
\end{table}

\begin{figure}[t]
\begin{center}
\includegraphics[scale=1.1]{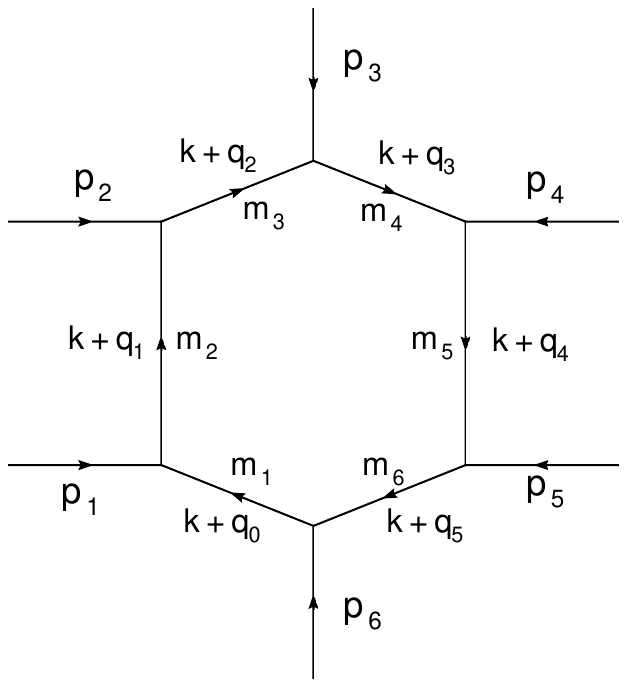}
\caption{\label{fig:6ptfig}
Momenta flow for the massive six-point topology.}
\end{center}
\end{figure}

As a second example, we reproduce components of the massive tensor integral $I_6^{\alpha\beta\gamma\delta\epsilon}$ in table \ref{sixpoint-massive}.
The kinematics is defined by figure \ref{fig:6ptfig} (with $q_0=0$) and table \ref{kinem-hexagon-massive}.
All tensor components  are finite.
The numbers could not be checked by another open source program, but we had the opportunity to compare them with an unpublished numerical package \cite{Uwer-privcommun:2009}.

\begin{table}[th]
\centering
\begin{tabular}{|r|r|r@{.}l|r@{.}l|r|}
\hline
$p_1$ & 0.21774554~E+03    &     0&0              & 0&0              & 0.21774554~E+03 \\
$p_2$ & 0.21774554~E+03    &     0&0              & 0&0              &  -- 0.21774554~E+03\\
$p_3$ & -- 0.20369415~E+03 &  -- 0&47579512~E+02  &0&42126823~E+02   &0.84097181~E+02   \\
$p_4$ & -- 0.20907237~E+03 &     0&55215961~E+02  &-- 0&46692034~E+02&  -- 0.90010087~E+02 \\
$p_5$ &  -- 0.68463308~E+01&     0&53063195~E+01  &  0&29698267~E+01 & -- 0.31456871~E+01  \\
$p_6$ & -- 0.15878244~E+02 &  -- 0&12942769~E+02  & 0&15953850~E+01  &  0.90585932~E+01
\\
\hline
\multicolumn{7}{|c|}{$m_1 = m_2 = m_3 = m_5 = m_6 = 110.0, ~~m_4 = 140.0$}
\\
\hline
\end{tabular}
\caption{\label{kinem-hexagon-massive}Randomly chosen phase space point of  six-point functions with massive particles.}
\end{table}

\begin{table}[htb]
\begin{center}
\begin{tabular}{|c|r@{.}l r@{.}l|}
\hline
  & \multicolumn{4}{|c|}{Hexagon.F}
\\
\hline
${F^{03121}}$  &  0&29834730E--09, & -- 0&68229122E--10
\\
\hline
${F^{11020}}$  &  0&42830755E--09, & 0&42574811E--09
\\
\hline
${F^{20200}}$  & -- 0&71172947E--08, & 0&10102923E--07
\\
\hline
${F^{22130}}$  & -- 0&29200434E--09, & 0&78553811E--10
\\
\hline
${F^{33333}}$  &   0&17451484E--07, & -- 0&30914316E--07
\\
 \hline
\end{tabular}
\end{center}
\caption[]{\label{sixpoint-massive}Selected tensor components of rank $R=5$ massive hexagon integrals;
}
\end{table}

Further  numerical results may be found in the transparencies of the talk \cite{Riemann-radcor:2009}.

\section*{Acknowledgments}
Work supported in part by Sonderforschungsbereich/Trans\-re\-gio SFB/TRR 9 of DFG
``Com\-pu\-ter\-ge\-st\"utz\-te Theoretische Teil\-chen\-phy\-sik"
and by the European Community's Marie-Curie Research Trai\-ning Network
MRTN-CT-2006-035505
``HEPTOOLS''.
J.F. likes to thank DESY for kind hospitality.

\providecommand{\href}[2]{#2}\begingroup\raggedright\endgroup

\end{document}